\documentclass[reqno]{article}
\usepackage{amsmath,amsfonts,amssymb,amsthm,hyperref,tabularx}

\renewcommand\Pr{\mathbb{P}}
\newcommand\E{\mathbb{E}}
\newcommand\Var{\mathrm{Var}}
\newcommand\In{\mathcal{I}}
\newcommand\Bi{\mathcal{B}}

\renewcommand\epsilon{\varepsilon}

\newcommand\bN{\mathbf{N}}
\newcommand\bT{\mathbf{T}}
\newcommand\bj{\mathbf{j}}
\newcommand\bJ{\mathbf{J}}
\newcommand\bn{\mathbf{n}}
\newcommand\bp{\mathbf{p}}

\newcommand\dx{\mathrm{d}x}
\newcommand\dy{\mathrm{d}y}
\newcommand\dt{\mathrm{d}t}
\newcommand\du{\mathrm{d}u}
\newcommand\defint{\int_{-\infty}^\infty}
\newcommand\pmin{p_{\mathrm{min}}}
\newcommand\est{\widehat{\delta t}}
\newcommand\hatp{\widehat p}
\renewcommand\le{\leqslant}
\renewcommand\ge{\geqslant}

\DeclareMathOperator*{\argmax}{arg\,max}
\begin{document}

\title{Statistical model uncertainty and OPERA-like time-of-flight measurements}
\author{Oliver Riordan%
\thanks{Mathematical Institute, University of Oxford, 24--29 St Giles', Oxford OX1 3LB, UK}
\and Alex Selby%
\thanks{alex.selby@cantab.net}
}
\date{1 November, 2011}
\maketitle

\begin{abstract}
Time-of-flight measurements such as the OPERA and MINOS experiments
rely crucially on statistical analysis (as well as many other
ingredients) for their conclusions. The nature of these experiments
leads to a simple class of statistical models for the results;
however, which model in the class is appropriate is not known exactly,
as this depends on information obtained experimentally, which is subject
to noise and other errors. To obtain robust conclusions, this problem,
known as ``model uncertainty,'' needs to be addressed, with quantitative
bounds on the effect such uncertainty may have on the final result.

The OPERA (and MINOS) analysis appears to take steps to mitigate the
effects of model uncertainty, though without quantifying any remaining
effect. We describe one of the strategies used (averaging individual
probability distributions), and point out a potential source of error
if this is not carried out correctly.  We then argue that the correct
version of this strategy is not the most effective, and suggest
possible alternatives. These alternatives may give more
accurate statistical results using the same data, allowing, for
example, more accurate determination of the dependence of the
anomalous time shift on energy.  Which strategies work and how well
can only be evaluated with access to the full data.

Whether or not the anomalous result from OPERA turns out to be confirmed,
we believe that techniques such as those presented here may be appropriate
for the analysis of other timing experiments of this type.
\end{abstract}

\section{Introduction}\label{intro}

The fundamental statistical question arising from the OPERA
timing experiment has the following form: the observations (neutrino arrival times) can be
seen as independent samples $T_i$ from a family $n_i(t)$ of probability
distributions shifted by a common unknown offset, and the
task is to estimate this offset. Unfortunately, the $n_i(t)$ are not known
exactly; rather some proxy measurement $p_i(t)$ is available, and the statistics must
take this into account. There may also be small errors in the measurement of the $T_i$
that are not independent.

In the specific case of the OPERA experiment, $p_i(t)$, the $i^{\rm th}$ proton waveform,
is the output of the waveform digitizer connected to the BCT (Beam Current Transformer)
for the $i^{\rm th}$ proton pulse (or ``spill''). Following \cite{OPERA}, we write $\delta
t$ for the (unknown) difference $\mathrm{TOF}_c-\mathrm{TOF}_\nu$\footnote{We follow the sign convention for $\delta t$ in~\cite{OPERA}.} between the actual
time-of-flight and that expected (distance divided by $c$).
Let
\[ 
 \Delta t=\mathrm{TOF}_c+t_{\nu\mathrm{md}}-t_{p\mathrm{md}}
\]
denote the expected time delay between the proton waveforms and neutrino
detection events in the case that neutrinos move at the speed of light,
where $t_{\nu\mathrm{md}}$ and $t_{p\mathrm{md}}$ incorporate all known neutrino and proton
measurement delays respectively.
Then we may define $n_i(t)$ by 
saying that detection events for the interval $[t+\Delta t-\delta t,t+\Delta t-\delta t+\dt)$ correspond to
a Poisson process with rate $n_i(t)\dt$.
The notation is set so that $p_i(t)$ and $n_i(t)$ have
the same time origin (and would be proportional under idealized conditions), but $n_i(t)$
is defined in terms of detection events. This gives us the notational
freedom to hypothesize any relationship between $p_i(t)$ and $n_i(t)$.  In this paper
we shall indicate how one may cope with certain unknown differences between them.

If the chance of a neutrino giving rise to a detection event is independent of its
position in the $10.5\mu$s pulse, then $n_i(t)$ can be thought of as the neutrino waveform.
This is a fairly mild assumption, but even so requires justification since, for example, it
could be imagined that for some reason the neutrinos towards the end of the $10.5\mu$s
pulse tend to be of slightly higher or lower energy than those at the start, and
that the detector has different sensitivities to neutrinos of different energies.

In the idealized case that the BCT measures the proton current exactly, that each
proton has the same chance of ultimately leading to a neutrino detection, and that
the relevant clocks are perfectly synchronized, then $n_i(t)$
and $p_i(t)$ are equal. However, they are clearly not exactly equal for a number of reasons;
see, for example, the discussion of oscillations on page 125 of~\cite{Brunetti}.
Since only $p_i(t)$ can be measured, but naive statistical analysis requires
knowledge of $n_i(t)$, a problem called ``model uncertainty'' arises.
To obtain reliable statistical results, it is necessary to quantify
the extent of this uncertainty (the difference between the $p_i(t)$ and the $n_i(t)$),
or at least to argue explicitly that it is negligible; it is \emph{also} necessary to quantify
how sensitive the output of the statistical estimation procedure
is to changes in the ``input'' $n_i(t)$.

More explicitly, given $n_i(t)$,
the task of estimating $\delta t$ is a simple closed mathematical problem.
We can think of the functions $n_i(t)$ as leading to a one-parameter
family (parameterized by $\delta t$) of models of when we expect neutrinos to be
detected, and then standard techniques such as maximum likelihood estimation will work well.
But in practice what we actually know is $p_i(t)$, so we don't know which model
$n_i(t)$ to use. As far as possible we'd like
to model the model uncertainty statistically and just make a bigger model, but some types
of uncertainty may be hard to model this way, for example because they might be non-independent
for different $i$.

In the arXiv preprint~\cite{OPERA} posted by
the OPERA collaboration it is implicitly assumed that any failure of
$n_i(t)$ to be proportional to $p_i(t)$ is accounted for in the systematic
errors. However, it is possible that discrepancies between $p_i(t)$ and $n_i(t)$ could
reinforce each other in such a way as to create a much greater discrepancy in the final
estimate of $\delta t$. Similarly, the estimate of $\delta t$ could be particularly sensitive
to small discrepancies in particular parts of the waveform; a qualitative
discussion of a potential effect of this type is given in~\cite{Knobloch}.
Therefore we suggest that the statistical analysis should be conducted so
as to be as robust as possible against such discrepancies of $n_i(t)$ from
$p_i(t)$, and we give examples of how this might be done for certain classes of
discrepancies.

One strategy adopted in~\cite{OPERA} for (it appears) precisely the purpose of reducing
the effect of noise in the measured current $p_i(t)$ involves
summing the $p_i(t)$ to make an aggregate proton waveform
$w(t)$. [For simplicity of notation, and because it is not relevant to the discussion here, we
omit the distinction between the first and second extractions; when we speak of summing
all proton waveforms, this can be understood to mean all waveforms within the first (or second)
extraction.]
We believe that this process is potentially flawed, for a reason given below, and if not
flawed, then it appears not to be making best use of the information available. If the
$p_i(t)$ are treated separately for each $i$, then it is likely that a more accurate result can
be obtained, which would, for example, enable one to probe the energy dependence of
$\delta t$ with greater precision. Whether this is in fact possible depends
on details of the waveforms $p_i(t)$, which do not seem to be publicly available
at the time of writing, as well as on properties of the discrepancies between the $n_i(t)$
and the $p_i(t)$.

\section{Summing proton waveforms}\label{summing}

In Section 7 of \cite{OPERA} and Section 8.1 of~\cite{Brunetti},
it is stated that the aggregate distribution (PDF) $w(t)$
is constructed by adding $p_i(t)$ over $i$ corresponding to neutrino detections, then
normalizing the result. This procedure is statistically flawed for reasons given below,
and if it had been used could very easily explain the anomalous result $\delta t>0$.
We raised this issue with one of the authors of~\cite{OPERA} (Dario Autiero), who informed
us in a private communication~\cite{Autiero} that
the $p_i(t)$ were in fact first normalized before being added together.
This avoids the above-mentioned flaw, but we believe that it may still
be useful to describe this effect, 
because the references \cite{OPERA,Brunetti} are not clear (or if anything,
clear the wrong way!) on this important point.

Treating (for the moment) $p_i(t)$ and $n_i(t)$ as interchangeable, the problem with
adding up the $p_i(t)$ is that it may be that $a_i=\defint p_i(t) \dt$ depends on $i$, i.e.,
that some pulses contain more protons that others. Only
those $p_i(t)$ leading to a (single, we assume) neutrino detection event are selected.
Conditioned on
$p_i(t)$ giving rise to a single neutrino event, the time of this
event (or rather, of the proton causing the event) is simply a random variable
with density given by the normalized version $\hatp_i(t)$ of $p_i(t)$.
Hence the statistically proper thing to do would be
to first normalize each $p_i(t)$ for each $i$ corresponding to a neutrino event, and then
add up the resulting $\hatp_i(t)$, and then normalize again
(i.e., divide by the total number of detections)
to get the aggregate distribution $w(t)$.

Summing the $p_i(t)$ and then normalizing corresponds to taking a weighted
average of the correct distributions $\hatp_i(t)$, with each weighted by $a_i$.
This is known as `size-biasing'. If there is a correlation between $a_i$
and the shape of the waveform $p_i(t)$ (which seems very plausible given that the 
individual $p_i(t)$ seem to differ significantly), then the size-biased average $\tilde w(t)$
may differ from $w(t)$. It is quite possible that these averages may have a very similar
shape, but with a small time offset. This could lead to an error in the estimate
of $\delta t$ that would not be picked up by the tests described in~\cite{OPERA},
such as the Monte Carlo simulation, assuming these tests also used $\tilde w(t)$
rather than $w(t)$ or the individual $p_i(t)$.

Adding up the $p_i(t)$ before normalizing would make sense if
\emph{all} $p_i(t)$ were included, not just those corresponding to neutrino detection events
(or, at least, the subset of the $p_i(t)$ to be added together is chosen without regard
to whether or not a neutrino was detected). This would construct a different
unbiased estimator of the aggregate distribution. It may be counterintuitive
that simply ignoring waveforms that did not lead to a detection can introduce
an error. However, taking this point
of view, selecting events leading to a detection introduces size-biasing, which
must then be reversed by normalizing.  We believe that summing all $p_i(t)$
is unlikely to be the best way to proceed statistically,
since it throws away information about which
neutrino events occurred, so from now on we assume that $i$ indexes only those pulses for
which a neutrino was detected.

\section{Using individual proton waveforms}\label{sec3}

From now on we assume that all individual waveforms $p_i(t)$ and $n_i(t)$ are normalized.
We ignore the known time offset $\Delta t$, assuming that the data $T_i$ now
represent detection times from which $\Delta t$ has been subtracted.
Later, we shall adopt the following model:
independent observations $N_1,N_2,\ldots,N_N$ are drawn from
$n_1(t),n_2(t),\ldots,n_N(t)$. The measured arrival times $T_1,\dots,T_N$
are then given by $T_i=N_i-\delta t+J_i$, where $\delta t$ is the parameter we
wish to measure and the $J_i$ represent small noise in the form of ``jitter''.
The actual observed values are $t_1,t_2,\ldots,t_N$, as in~\cite[Section 7]{OPERA}, where
$N=16111$, but we use $T_1,T_2,\ldots,T_N$ when we want to think of these observations as
random variables.

For the moment, assume that $J_i=0$ and $p_i(t)=n_i(t)$. It is then a purely mathematical
problem to estimate $\delta t$, since we have (temporarily) removed all sources of errors
in this description. A good method to use here is maximum likelihood estimation: if
$\delta$ is a candidate value for the unknown shift, then writing $\bT$ for the
collection $(T_1,T_2,\ldots,T_N)$, the log likelihood is
\begin{equation}\label{LL}
  L(\delta;\bT)=\sum_{i=1}^N \log(p_i(T_i+\delta))
\end{equation}
and the maximum likelihood estimator (MLE) of $\delta t$ is $\est(\bT)$, the value of
$\delta$ that maximizes $L(\delta;\bT)$.

Intuitively, most of the statistical benefit of a good estimator for $\delta t$ comes from
the parts of the input waveforms, $p_i(t)$, that are varying the fastest. We can quantify
this by taking expectations over $T_i$ to look at typical values of $L(\delta;\bT)$: the
expectation is
\[
  \E L(\delta)=\sum_i \defint p_i(t+\delta t)\log(p_i(t+\delta))\,\dt.
\]

Differentiating with respect to $\delta$ and substituting $t$ for $t+\delta t$ in the integral, we see that
\begin{equation}
  \E L'(\delta t)=\sum_i\defint p_i(t)\frac{p_i'(t)}{p_i(t)}\dt=0,\label{bias}
\end{equation}
which means that the MLE will be asymptotically unbiased.
$-\E L''(\delta t)$ (the Fisher information) is a measure of how responsive the likelihood is
to changes in $\delta$, and so quantifies the information available. It can be identified with
the expected second derivative of the graphs in \cite[Fig. 8]{OPERA} at the true value
$\delta t$ (which may be slightly different from the second derivative at the maximum of
these graphs).

The MLE asymptotically achieves this Fisher information, which has an expression
independent of the unknown $\delta t$:
\begin{equation}
  -\E L''(\delta t)=\sum_i\In(p_i())=\sum_i\defint\frac{p_i'(t)^2}{p_i(t)}\dt\label{fisher}.
\end{equation}
(More precisely, assuming the integral converges, then if the distribution or family
of distributions is fixed and the number of samples tends to infinity, then the
standard deviation of $\est$ will scale with $(-\E L''(\delta t))^{-1/2}$.)

As expected, the more wiggly the waveforms are, the better, especially if the wiggle is where $p_i(t)$ is
small. In practice, there are limits to how much use can be made of the
regions where $p_i(t)$ is small, due to small uncertainties in the value of $p_i(t)$,
and to the limited number of samples available; the latter effect means that the asymptotic
regime is not necessarily attained.

In contrast, treating the samples as independent samples from a single aggregate
distribution $w(t)=N^{-1}\sum_{i=1}^Np_i(t)$, our estimator is the maximum of the log
likelihood in~\eqref{LL} with $w()$ in place of $p_i()$.  The information available by
this method is $N\In(w())$, and it is easy to check that $N\In(w()) \le \sum_{i=1}^N
\In(p_i())$, with equality if and only if all the $p_i(t)$ are equal.  Thus averaging the
$p_i(t)$ will result in a less sensitive (higher variance) estimator.

The interpretation of this is that, as might be expected, the procedure described
in~\cite{OPERA} of averaging together all the proton distributions will wash away
information that is present in the individual distributions. The amount of information
lost will depend on how different the individual distributions are, but comparing the
graphs in Figures 4 and 9 of~\cite{OPERA} it seems that these individual distributions are
indeed very different, and much more wiggly than the aggregate distribution. This suggests
that an estimator of $\delta t$ based on individual distributions could be considerably
more accurate than the existing estimator. (Even if it is not possible to effectively
take advantage of the fine structure of the $p_i$, it should be possible
to benefit from the 5-peak structure much more strongly present in Figure 4 than in Figure 9.)

The above argument assumed that $n_i(t)=p_i(t)$ and $J_i=0$ (i.e., $T_i=N_i-\delta t$):
to be correct it should have been written
in terms of $N_i-\delta t$ and $n_i(t)$, not $T_i$ and $p_i(t)$.
Since $J_i$ and $n_i(t)$ are unknown, we may be tempted to
forget about these differences, but unfortunately the
estimator $\est(\bT)$ so obtained might be quite susceptible to small discrepancies between
$p_i(t)$ and $n_i(t)$, or to jitter $J_i$.
For example, for a particular $i$, we might have $p_i(t)=0$ for $t$
near to $t_i+\delta t$, which would cause it to be impossible to estimate $\delta t$ near
its true value, regardless of any amount of other data. 

According to a remark in a private communication~\cite{Autiero}, this kind of effect
(specifically ``small white noise (electronics) on the baseline'') in each $p_i(t)$
is the reason why individual $p_i(t)$ were not used in~\cite{OPERA}.
However, we believe that it should be possible to construct estimators that are
robust against such effects, and therefore enable us to tap into the large amount
of extra information available from the individual $p_i(t)$. This is the subject
of the next section.

Note that averaging the $p_i(t)$ does not completely eliminate noise;
it merely reduces its variance. When using statistics to extract a very delicate
result (order of $10$ns accuracy) from a wide spread (approx $10\mu$s), the small residual
discrepancy between $\sum n_i(t)$ and $\sum p_i(t)$ may yet be important. In other words,
whatever method is chosen (averaging $p_i(t)$ or not), it is still necessary to have a
robust statistical appraisal of the effects of possible discrepancies between $p_i(t)$ and
$n_i(t)$.

It seems then that a proper statement of a statistical analysis of this type
should always be preceded by a caveat of
the form ``Assuming that $p_i(t)$ and $n_i(t)$ can differ in the following sorts of ways,
then...''.  This is the limit of what can be said without further experimental data, since
if we have no bound at all on how different $p_i(t)$ and $n_i(t)$ can be, then obviously
no useful information can be extracted from any statistics.

\section{Constructing robust estimators}

\subsection{General procedure}\label{ss_error}

In accordance with the above prescription, we first decide on the class of discrepancies
between the actual distribution of $\bT=(T_1,\ldots,T_N)$ and its idealized version
that we shall work with. Then we choose 
an estimator, and finally we measure the bias and efficiency of the estimator by
finding lower bounds on the probabilities of confidence intervals centred on the estimator.

We shall assume that
\[
  p_i(t)=n_i(t)+\epsilon_i(t),
\]
where
\begin{equation}\label{TVbound}
 \defint |\epsilon_i(t)|\dt\le 2\epsilon,
\end{equation}
corresponding to a total variation distance of at most $\epsilon$.
We write $p_i\approx n_i$ whenever this holds, and $\bp\approx\bn$ to indicate
that it holds for every $i$.

We assume that $\bN=(N_1,\ldots,N_N)$ consists of independent samples from
the probability densities $n_1(t),\ldots,n_N(t)$, and that
\[
 T_i = N_i-\delta t +J_i
\]
where
\begin{equation}\label{jitter}
 |J_i|\le J
\end{equation}
for each $i$.

We think of $\epsilon_i(t)$ as representing ``noise'' in the measurement $p_i$
of $n_i$, and the ``jitter'' $J_i$ as including errors in the measurement
of the arrival times, and so on. There is some redundancy here, since
certain types of difference between the distributions $p_i$ and $n_i$
can also be seen as jitter. 
Note that these
discrepancies are worse than statistical (independent random) or systematic (random, but
constant over $i$) error, since they are assumed to be chosen non-randomly ---
``maliciously'' --- in such a way as to change the estimator from the true value by as
much as possible. The motivation for this is that we may not be sure which errors are
independent, and we wish to avoid having to describe all possible
ways in which they might be correlated. For example, we should like our estimator to be
robust against small timing errors that are somehow correlated with the time offset of
the proton within the pulse, or timing errors that result from very slight unknown
changes in distance and hence $\mathrm{TOF}_c$ over the months and years.

If there were no jitter, then the 200MHz signal present in each $p_i(t)$ (\cite[pp.\ 5--6]{OPERA})
would be extremely useful in pinning down a good estimate of $\delta t$. But to
be on the safe side we presume that there is jitter of the order of at least 2.5ns, which
means that the 200MHz information is illusory, and in fact we must be careful not to try
to make use of it, since +/- 2.5ns jitter could cause disproportionately large changes in
the likelihood by means of aligning, or anti-aligning, the peaks of the 200MHz signal in
$p_i(t)$ with the data.
This argument illustrates the kind of effect we need to be robust against.

If we knew the distribution of $\bT$, then we could define any
estimator $\est(\bT)$, and establish statistical bounds on
$|\est(\bT)-\delta t|$ (i.e., a confidence interval) simply by
repeatedly generating samples from the (vector valued) distribution
$\bT$ and evaluating the difference $|\est(\bT)-\delta t|$.  (Since
$\delta t$ is unknown, in general one has to consider all possible
values of $\delta t$, but as we shall see, this is not necessary here.)
Of course, we do not know this distribution exactly.

Our assumptions imply that we can write $T_i=N_i-\delta t+J_i$,
where the $N_i$ are independent with distributions $n_i(t)$
satisfying $n_i\approx p_i$.
Note that we make no independence assumptions concerning the $J_i$ or $T_i$.
If we can show that
\begin{equation}
  \forall \bn\approx \bp: \Pr\left( \forall |J_i|\le J: |\est(\bT)-\delta t|\le w\right)\ge s\label{conf}
\end{equation}
then the estimator $\est(\bT)$ can be said to be within $w$ of $\delta t$ with a
confidence of $s$, whatever the unknown $n_i(t)$ and $J_i$ are.\footnote{ For simplicity
  we intially consider symmetric confidence intervals, and indicate a suitable
  modification for asymmetric confidence intervals at the end of this section. This may be
  more appropriate if the final estimator has a small asymptotic bias.}
For this to be useful
we need two properties: (i) that we haven't given away too much, i.e., that values of $w$ and
$s$ satisfying (\ref{conf}) give reasonably small error bounds,
and (ii) there is an effective way to show that (\ref{conf}) is satisfied for given $w$
and $s$, since, for example, simply exhausting over the space $[-J,J]^N$ of all possible
jitters is computationally impractical. Fortunately, it is possible to satisfy
both requirements simultaneously.

\subsection{Choice of estimator}\label{estchoice}

The estimators we shall consider here can be thought of as modified maximum
likelihood estimators: we choose a modified log likelihood, $f_i(t)$, then,
analogously to (\ref{LL}), define $\est(\bT)$ to
be the value of $\delta$ that maximizes the modified likelihood:
\begin{equation}\begin{split}
  \tilde L(\delta;\bT)&=\sum_i f_i(T_i+\delta),\\
  \est(\bT)&=\argmax_\delta(\tilde L(\delta;\bT)).
\end{split}\end{equation}

Although the method we describe is mathematically rigorous for any choice of $f_i(t)$,
the results (width of the final interval) will vary depending on the choice of $f_i(t)$,
which must be adapted to the real data $p_i(t)$ and assumptions made
in place of \eqref{TVbound} and \eqref{jitter}.

We shall consider $f_i(t)$ defined by
\begin{equation}\label{modll}
  f_i(t)=\defint\log(\max\{p_i(u),\pmin\})\phi_\alpha(t-u)\du,
\end{equation}
where
\[
  \phi_\alpha(x)=\frac{1}{\alpha\sqrt{2\pi}}\exp(-x^2/(2\alpha^2))
\]
is a Gaussian with width $\alpha$, and $\pmin$ and $\alpha$ are suitably chosen.
(Convolving $p_i(t)$ with $\phi_\alpha(t)$ before
taking the logarithm also appears to work reasonably well, though convolving after taking
the logarithm, as here, is slightly easier to reason about.)

Before turning to the verification that such an $f_i(t)$ gives good
results (narrow confidence intervals), we briefly 
motivate this choice for $f_i(t)$. 

Setting a minimum probability level, $\pmin$, will greatly reduce sensitivity
to noise where $p_i(t)$ is small, and hence $\log(p_i(t))$ is large in magnitude.
This change will not affect the asymptotic bias in the idealized case (when $n_i=p_i$ and $J_i=0$).
Indeed, in place of (\ref{bias}) 
we have the modified bias formula
\begin{equation}
  \Bi(f_i(),p_i())=\sum_i \defint p_i(t)f_i'(t)\dt=-\sum_i \defint p_i'(t)f_i(t)\dt. \label{mbias}
\end{equation}
When $f_i(t)=\log(\max\{p_i(u),\pmin\})$, it is easy to check that $\Bi(f_i(),p_i())=0$, so
putting a floor on the value of $p_i(t)$ does not in itself significantly bias the estimator.
It does increase its variance; this is part of a trade-off between variance
of the estimator as applicable in ideal conditions, and sensitivity to jitter and noise.

The other modification in (\ref{modll}) is convolution with the Gaussian $\phi_\alpha$;
for the OPERA data this will remove the 200MHz signal if $\alpha\gg 5$ns, and
can be expected to prevent jitter from disproportionately affecting the likelihood
if $\alpha\gg J$. There is an effect on the bias, but it is only of order $O(\alpha^2)$
as $\alpha\to0$ since $\phi_\alpha(t)$ is symmetric under $t\leftrightarrow-t$.

\subsection{Analysis of significance of estimator}\label{anasig}

The unknown parameter $\delta t$ appears in \eqref{conf}, but 
can easily be eliminated. Indeed,
the estimator we use is translation invariant, in that
\[
 \est(T_1+\tau,T_2+\tau,\ldots,T_N+\tau)=\est(T_1,T_2,\ldots,T_N)-\tau.
\]
(The minus sign here comes from the following the sign convention for $\delta t$
in~\cite{OPERA}.) The difference $\est(\bT)-\delta t$ that we are interested in can thus be written
as $\est(\tilde\bT)$, where $\tilde T_i=T_i+\delta t=N_i+J_i$. Let
\begin{equation}\label{conf3}\begin{split}
 L(\delta;\tilde \bT)=L(\delta;\bN,\bJ)&=\sum_i f_i(N_i+J_i+\delta),\\
  m(\bN,\bJ)&=\argmax_\delta(L(\delta;\bN,\bJ)),\\
  s_1(w)&= \inf \Pr(|m(\bN,\bj)|\le w)\\
&=\inf_{\bn\approx \bp} \Pr\bigl( \forall\,\bj\in [-J,J]^N: |m(\bN,\bj)|\le w\bigr).
\end{split}\end{equation}
In the second last line, the infimum is over all allowed distributions of $\tilde\bT=\bN+\bJ$;
the final expression follows since we make no assumptions
on $\bJ$ other than that $|J_i|\le J$ for each $i$, so the worst value $\bj$ of $\bJ$
can be chosen after seeing $\bN$. As discussed above, $\est(\bT)\pm w$ is a confidence
interval for $\delta t$ with confidence level $s_1(w)$.

Unfortunately, the space of possible values of the $n_i$ (depending on the noise)
and the space $[-J,J]^N$ of jitter values are high-dimensional.
Because the position of the maximum of $L(\delta)$ can theoretically change a lot given a
small change in one of the summands $f_i(N_i+J_i+\delta)$, optimizing over these
spaces is likely to be computationally intractable. To cope with this, we replace $s_1(w)$ with a
possibly slightly weaker (though in practice very similar) estimate based on derivatives.

To simplify the exposition slightly, we assume \emph{a priori} that the maximum
of $L(\delta)$ will occur in a limited range
$[-w_\mathrm{max},w_\mathrm{max}]$. (In practice, the probability
of this assumption failing can be estimated by techniques
similar to those described below.)
We use the observation that, in this case, if the derivative $L'$
is positive on $[-w_\mathrm{max},-w]$ and negative on $[w,w_\mathrm{max}]$,
then the maximum over $[-w_\mathrm{max},w_\mathrm{max}]$ lies in $[-w,w]$. The point is that we can
express the derivative of $L$ as a sum of individual derivatives,
and bound them separately. Thus we set
\begin{equation}\label{conf4}\begin{split}
  D_\mathrm{min}(\delta,\bN)&=\inf_{\bj}L'(\delta;\bN,\bj),\\
  D_\mathrm{max}(\delta,\bN)&=\sup_{\bj}L'(\delta;\bN,\bj),\\
  s_2(w)&=\inf_{\bn\approx \bp}\ \Pr\Bigl(\forall\delta\in[-w_\mathrm{max},-w]\; D_\mathrm{min}(\delta,\bN)>0\; \mathrm{and}\\
      & \hskip1.85cm\forall\delta\in[w,w_\mathrm{max}]\; D_\mathrm{max}(\delta,\bN)<0\Bigr),
\end{split}\end{equation}
where the infimum is over all allowed distributions of $\bN$, i.e., over all possible values of the noise.

This ensures that whatever the unknown noise and jitter,
with probability at least $s_2(w)$ the function $L(\delta;\tilde \bT)=L(\delta;\bN,\bJ)$
will be increasing on $[-w_\mathrm{max},-w]$ and decreasing on $[w,w_\mathrm{max}]$, and so
will have a maximum in $[-w,w]$ that is global over $[-w_\mathrm{max},w_\mathrm{max}]$.
This implies that $|m(\bN,\bJ)|\le w$, assuming that $L(\delta;\bN,\bJ)$ does not have a rival maximum
further away than $w_\mathrm{max}$.

The use of $w_\mathrm{max}$ is a technicality, and can be removed with
a slightly refined version of this argument, but we retain it for
simplicity.  (Alternatively, it happens that for our approximate
reconstruction of the OPERA data, even choosing $w_\mathrm{max}$ as
large as $250$ns has only a negligible effect on our final confidence
bound ($s_3(w)$; see below). But the interval
$[-250\mathrm{ns},250\mathrm{ns}]$ is wide enough to encompass all
plausible values of $\delta t$.)

The advantage of calculating a version of $s(w)$ in terms of the derivative of $L$ as in
(\ref{conf4}), is that the optimization over $\bj$ becomes tractable: it decomposes into
separate optimizations over each $j_i$. For example,
\begin{align}
  D_\mathrm{min}(\delta,\bN)&=\inf_{\bj}L'(\delta;\bN,\bj) \nonumber \\
                             &=\sum_i\inf_{j_i}f_i'(N_i+j_i+\delta).\label{dmsum}
\end{align}

There does not seem to be a simple way to handle the final infimum in \eqref{conf4},
but we may replace it with a sum of more easily calculated terms while only modestly
weakening the confidence bound. From now on we discretize time,
by rounding all $T_i$ to multiples of some small constant, say $1$ns. The effect
of any errors this introduces is bounded by increasing $J$ by $0.5$ns. All
derivatives then become simply differences of consecutive values.
For $\delta\in [-w_{\mathrm{max}},-w]$ let $F_\delta$ be the ``failure'' event that $D_\mathrm{min}(\delta,\bN)\le 0$,
and for $\delta\in [w,w_{\mathrm{max}}]$ let $F_\delta$ be the event that $D_\mathrm{max}(\delta,\bN)\ge 0$.
Then $s_2(w)$ as defined above is the infimum of the probability that no $F_\delta$ holds. Hence
\begin{equation}\begin{split}
 s_2(w) &\ge \inf_{\bn\approx\bp} \biggl\{1- \sum_{w\le|\delta|\le w_{\mathrm{max}}} \Pr(F_\delta) \biggr\} \\
 &\ge 1- \sum_{w\le|\delta|\le w_{\mathrm{max}}} \sup_{\bn\approx\bp} \Pr(F_\delta) \ =\  s_3(w).
\end{split}\end{equation}
The final supremum is easy to calculate. Suppose, for example,
that $\delta>0$. Then $F_\delta$ is the event that $S_\delta=\sum_i g_{i,\delta}(N_i)\ge 0$,
where $g_{i,\delta}(x) = \sup_{|j|\le J} f_i'(x+j+\delta)$ is a function of $x$
that is easy to calculate from the known data.
The assumption $n_i\approx p_i$ implies that $N_i$ can be coupled with a variable
$X_i$ having the known density $p_i(t)$ so that $\Pr(X_i\ne N_i)\le \epsilon$.
If we choose the distribution of $N_i$ 
by removing the ``most helpful'' $\epsilon$-fraction of the distribution of $X_i$,
i.e., the part where $g_{i,\delta}(X_i)$ is smallest, and replacing
it by probability $\epsilon$ of having the least helpful value,
then the distribution of this individual summand $g_{i,\delta}(N_i)$
is ``worse than'' (stochastically dominates) any other possible distribution where
$n_i\approx p_i$. Since $S_\delta$ is a sum of independent terms, the worst case overall
is given by combining these individual worst case distributions. (Note that the
worst case will be different for different values of $\delta$.)

It remains to estimate the probability that this sum of $N$ independent variables has
the right sign. This can be done efficiently by the method described in Appendix 1, and a
short example program implementing this can be found associated with this preprint.  This
could also be done by Monte Carlo simulation, though obviously it would be quite slow to
estimate very low tail probabilities this way.

A slight modification will improve the confidence bound in the case that there is a small
bias in the estimator. Instead of summing over
$\delta\in[-w_\mathrm{max},-w]\cup[w,w_\mathrm{max}]$, we sum over
$\delta\in[-w_\mathrm{max},-w+b]\cup[w+b,w_\mathrm{max}]$, on the assumption that the
modified estimator is
\begin{equation}\label{est}
 \est(\bT)=\argmax_\delta(\tilde L(\delta;\bT))+b,
\end{equation}
where $b=b(w)$ is determined once and for all by optimizing the resulting confidence estimate:
\[
 s_4(w) = 1 - \min_{|b|\le w_\mathrm{max}-w}\left( \sum_{\substack{\delta\in[-w_\mathrm{max},-w+b]\cup \\ [w+b,w_\mathrm{max}]}} \sup_{\bn\approx\bp} \Pr(F_\delta)\right).
\]

\section{Performance of estimator}

In order to properly evaluate the performance of the $\est(\bT)$ as defined by
(\ref{est}), we would need to know the 16111 proton waveforms, $p_i(t)$. These do not
appear to be publicly available at the time of writing, so instead we use the $p_i(t)$
given by \cite[Fig. 4]{OPERA} and, on the assumption that it is typical, set all $p_i$
equal to this one. This graph, whose axes are not displayed, was interpreted on the
assumption that the vertical axis is a linear scale with an offset. The zero point was
assumed to be at the mean of the left and right tail values, which are assumed to be noise.
After that, a plausible 200MHz signal was added.

In addition to the $p_i(t)$, we would need to know what assumptions can be
made about the various sources of error. Again, these are not described in~\cite{OPERA},
so we use the error model described in Section~\ref{ss_error}
with $0\le J\le3$(ns) and $\epsilon=0$, $10^{-4}$ or $10^{-3}$. Then the estimator parameters,
$\alpha$ and $p_\mathrm{min}$, were chosen to optimize the confidence for a 30ns error
($w=30$).

In practice, the amount of smoothing chosen was always large enough ($\alpha\ge22$) to
filter out almost all of any 200MHz signal. We might hope that the majority of the
statistical noise would also be eliminated by such smoothing, leaving $\epsilon=10^{-4}$
to account for any remaining ``malicious'' noise.

\begin{table}
\begin{center}
\begin{tabular}{r r r r r r}
   $J$ & $\epsilon$ & $\alpha$ & $p'_\mathrm{min}$ & $1-s_4(30)$ & $\sigma$\\
\hline
  0 &     0      &   22  &  $10^{-5}$          &  $5.0\times10^{-13}$ & 7.2 \\
  0 &  $10^{-4}$ &   29  &  $7\times10^{-6}$   &  $4.6\times10^{-11}$ & 6.6 \\ 
  0 &  $10^{-3}$ &   42  &  $3\times10^{-3}$   &  $4.4\times10^{-5}$  & 4.1 \\ 
  1 &     0      &   40  &  $10^{-8}$          &  $4.0\times10^{-10}$ & 6.3 \\
  1 &  $10^{-4}$ &   44  &  $10^{-5}$          &  $8.5\times10^{-9}$  & 5.8 \\
  1 &  $10^{-3}$ &   70  &  $1.7\times10^{-3}$ &  $9.4\times10^{-4}$  & 3.3 \\
  2 &     0      &   46  &  $10^{-7}$          &  $2.8\times10^{-8}$  & 5.5 \\
  2 &  $10^{-4}$ &   55  &  $10^{-6}$          &  $2.7\times10^{-7}$  & 5.1 \\
  2 &  $10^{-3}$ &   88  &  $9\times10^{-4}$   &  $6.5\times10^{-3}$  & 2.7 \\
  3 &     0      &   60  &  $10^{-9}$          &  $6.0\times10^{-7}$  & 5.0 \\
  3 &  $10^{-4}$ &   58  &  $10^{-7}$          &  $4.2\times10^{-6}$  & 4.6 \\
  3 &  $10^{-3}$ &  110  &  $7\times10^{-4}$   &  $2.8\times10^{-2}$  & 2.2 \\
\hline
  \multicolumn{4}{c}{OPERA COMPARISON} & $1.4\times10^{-5}$ & 4.3 \\
\end{tabular}
\end{center}

\caption{Tail confidence values $1-s_4(30)$, showing the confidence of the estimator being
  within 30ns of the true value under various different robustness
  scenarios. $p'_\mathrm{min}$ is linked to $p_\mathrm{min}$ by
  $\Pr(X_i<p_\mathrm{min})=p'_\mathrm{min}$. The last column shows the equivalent number
  of standard deviations for a Normal distribution: $s_4(30)$ is equal to the probability
  that a standard Normal is less than $\sigma$ in magnitude. The last row shows the tail
  confidence figure using the methods of \cite{OPERA} based on a 6.9ns statistical error.
}
\label{t1}

\end{table}

The results shown in Table~\ref{t1} can be compared with the statistical component of the
error as reported in \cite{OPERA}. We note that $6.9$ (stat.) from \cite[p.\ 19]{OPERA}
means that the confidence that the estimator from \cite{OPERA} is within 30ns of the true
value would be $1-\Phi(-30/6.9)/2=1-1.4\times10^{-5}$.  If we assume that $J=2$,
$\epsilon=10^{-4}$ is sufficiently pessimistic, then the corresponding tail confidence
using the methods of this paper would be $2.7\times10^{-7}$.  Alternatively, using
these methods, this same level of significance is reached for the same confidence bound of
30ns using only approximately $(4.3/5.1)^2=71\%$ (in fact 74\%) of
the current requirement of $N=16111$ observations. If $J=1$ is realistic, then
9800 observations (61\%) would be sufficient for the same confidence level.

These comparisons obviously rely on the assumption that we are using a typical $p_i$, and
the assumption that our jitter and noise values are appropriate, though we would expect
these methods to be useful to at least some extent in any case.  Note that we are
comparing a method with provable robustness against (the given type of) jitter and noise
with one where the effects of these are not quantified.  If the method of \cite{OPERA}
were adjusted to allow for these effects, the difference might be greater, although it is
plausible that there is little noise effect using the method of \cite{OPERA}, since that
is the benefit of adding up different $p_i$.
 
\section{Conclusions and discussion}

We have described what may be a more effective and more robust way of analysing the OPERA
neutrino time-of-flight data, and in general data from similar experimental set-ups. To do
this we have presented some statistical techniques together with an analysis of their
effectiveness.

It is only possible to properly gauge the effectiveness of these techniques when applied
to the OPERA data with access to the full experimental data, which does not seem to be
publicly available at the present time.

It is possible that there are types of differences between the proton waveforms,
$p_i(t)$, and the distribution of neutrino detection events other than the noise and
jitter we have considered here.  For example, in Section 8.1 of~\cite{Brunetti} it is
mentioned that there are spurious oscillations with periods $30$ns and $60$ns in some of
the $p_i(t)$. These are not mentioned in the more recent OPERA paper~\cite{OPERA}, but if
they are still present amongst the 16111 OPERA samples, then a new evaluation of estimator
effectiveness ought to be carried out, and possibly a new kind of estimator would be
required. It may be possible to use techniques similar to those of this paper to do
this, though it is not possible to say for sure without seeing a more detailed exposition
of the nature of these oscillations. In~\cite{Brunetti} it is explained that the spurious
oscillations are filtered out with an 8MHz low-pass filter;
the necessary evaluation of the effect that this may have on the estimator is
not described there. It may be
thought that the Monte Carlo techniques of~\cite{Brunetti} and~\cite{OPERA} would be
sufficient to catch a poor or biased estimator. However, these Monte Carlo
simulations appear to be based on samples from
the aggregate (summed) waveform distribution. This is not
correct in principle and will fail to catch some classes of problems. In addition, such
Monte Carlo methods lack the ability to simulate worst-case situations, as we
have sought to do in this paper in the case of noise and jitter.

\section{Appendix 1: Sums of independent variables}

We outline here the (fairly standard) method we used to bound $\Pr(S\ge 0)$,
where $S$ is a sum of independent variables $A_1,\ldots,A_N$ with $\E S\le 0$.
The method is based on Cram\'er's exponential tilting trick plus the Berry--Esseen
theorem. For notational
convenience we take the $A_i$ to have the same distribution $A$; the method
also works for different distributions by using a single tilting parameter $\theta$
chosen to achieve $\E T=\sum_i \E B_i=0$ below.

Let $a(x)$ denote the probability density function of $A$. By assumption
$\E A=\int xa(x)\dx\le0$. It follows (unless $A$ is never positive)
that there is some $\theta\ge0$ such that
$\E A e^{\theta A}=0$, i.e., such that $\int xe^{\theta x}a(x)\dx=0$.
Let $B$ denote the distribution with density function
\[
 b(x) = Z^{-1} e^{\theta x}a(x),
\]
where
\[
 Z= \int e^{\theta x}a(x)\dx
\]
is a normalizing constant. Then $\E[B]=0$.
Let $S$ denote the sum of $N$ independent variables with distribution $A$, and $T$
the sum of $N$ independent variables with distribution $B$. A standard
calculation shows that their densities are related by
\[
 t(x) = Z^{-N} e^{\theta x} s(x)  \quad\hbox{ and }\quad s(x)=Z^Ne^{-\theta x}t(x).
\]
It follows that
\[
 \Pr(S\ge 0) = \int_0^\infty Z^N e^{-\theta x}t(x)\dx.
\]

Let
\[
 T(x)=\Pr(T\ge x) = \int_x^\infty t(y)\dy.
\]
Then (integrating by parts) we can rewrite the formula above as
\[
 \Pr(S\ge 0) = Z^N T(0) - Z^N \int_0^\infty \theta e^{-\theta x} T(x) \dx.
\]
Of course, we do not know $T(x)$ in a practical way, but by the Berry--Esseen theorem
(with a modern form of the constant from \cite{Tyurin}), we have
\[
 |T(x)-U(x)|\le \eta = 0.4785 \frac{\E |B|^3}{(\Var B)^{3/2}} N^{-1/2}
\]
where $U(x)$ is the probability that a normal distribution with 
the same mean (0 here) and variance as $T$ exceeds $x$.
The worst case is given by taking $T(0)=U(0)+\eta=1/2+\eta$
and $T(x)=U(x)-\eta$ for $x>0$ such that $U(x)\ge \eta$,
and $T(x)=0$ for larger $x$. Writing $\tau=\theta\sigma$, where $\sigma^2$
is the variance of $T$, and substituting $x=\sigma y$ in the integral, this gives
\begin{align*}
 \Pr(S\ge 0) &\le Z^N (1/2+\eta) - Z^N \int_0^\infty \theta e^{-\theta x} \max\{U(x)-\eta,0\} \dx\\
             &\le Z^N (1/2+\eta) - Z^N \int_0^\infty \theta e^{-\theta x} (U(x)-\eta) \dx\\
             &= Z^N (1/2+2\eta) -  Z^N \int_0^\infty \theta e^{-\theta x} U(x) \dx\\
             &= Z^N (1/2+2\eta) -  Z^N \int_0^\infty \tau e^{-\tau y} (1-\Phi(y)) \dy\\
             &= Z^N(e^{\tau^2/2}\Phi(-\tau)+2\eta),
\end{align*}
where $\Phi$ denotes the distribution function of a standard Gaussian random variable.

\section{Appendix 2: Summary of notation}

\begin{tabularx}{\linewidth}{r X}\label{glossary}
  $p_i(t)$&The $i^{\mathrm{th}}$ distribution output by the waveform digitizer, with unspecified time origin,
normalized from Section~\ref{sec3} onwards.\\
  $n_i(t)$&The normalized $i^{\mathrm{th}}$ detection-event distribution, where the origin
           of time is that of $p_i(t)$, offset by the time-of-flight of neutrinos from CERN to
           OPERA, together with all known measurement delays. (See Introduction for details.)\\
  $\delta t$&The (unknown) early arrival time of the neutrinos at OPERA, i.e., 
             $\mathrm{TOF}_c-\mathrm{TOF}_\nu$.\\
  $\est(\bT)$&Estimator for $\delta t$.\\
  $N$&The number of neutrino detection events. Taken to be 16111.\\
  $t_i$&The measured observation times of neutrino detection events, in the same timeframe as
        that of $n_i(t)$.\\
  $T_i$&Arrival times in the statistical model: the $t_i$ are the actual measured values of the $T_i$.\\
  $\bT$&The collection $(T_1,\ldots,T_N)$.\\
  $\bN$&A sequence of independent random variables $N_1,\ldots,N_N$ with densities $n_i$.\\
  $f_i$&A function, derived from $p_i$, that will play the role of $\log p_i$ in a modified-log-likelihood estimator.\\
  $\delta$&In Section~\ref{estchoice}, $\delta$ is a candidate value for the unknown $\delta t$.
           In Section~\ref{anasig}, $\delta t$ has been subtracted out and $\delta=0$ is the true value.\\
  $J_i$&Small (actual, unknown) ``jitter'' defined by $T_i=N_i-\delta t+J_i$.\\
  $\bJ$&The collection $(J_1,\ldots,J_N)$.\\
  $j_i$&Candidate $J_i$ when optimizing over all possible jitter.\\
  $\bj$&The collection $(j_1,\ldots,j_N)$.\\
  $\epsilon$&Maximum noise considered (Section~\ref{ss_error}).\\
  $J$&Maximum jitter considered: $|J_i|\le J$ (Section~\ref{ss_error}).\\
  $\pmin$&Floor parameter used in defining modified MLE (Section~\ref{estchoice}).\\
  $\alpha$&Smoothing parameter used in defining modified MLE (Section~\ref{estchoice}).\\
  $s_i(w)$&($i=1,2,3,4$) Successively weaker, but more tractable/better, versions of the confidence
           bound: $\Pr(|\est(\bT)-\delta t|\le w)\ge s_i(w)$.\\
  $\Phi(x)$&$\Pr(X\le x)$ where $X\sim N(0,1)$.
\end{tabularx}

\end{document}